\documentclass[sigconf, screen, nonacm]{acmart}

\usepackage{fancyhdr}
\usepackage[]{hyperref}
\usepackage{subcaption}
\usepackage{color}
\usepackage[table,dvipsnames]{xcolor}
\usepackage{fancybox} 
\usepackage{makecell}
\usepackage[export]{adjustbox}
\usepackage[ruled, vlined, linesnumbered]{algorithm2e}
\usepackage{algpseudocode}
\usepackage{multirow, booktabs}
\usepackage{tikz}
\usepackage{listings}
\usepackage{bbding}
\usepackage{pifont}
\usepackage{amsthm} 
\usepackage{colortbl}
\usepackage{tabularx}
\usepackage{enumitem} 

\usepackage{soul}
\soulregister{\ref}7
\soulregister{\cite}7
\soulregister{\eqref}7

\def\BibTeX{{\rm B\kern-.05em{\sc i\kern-.025em b}\kern-.08em
    T\kern-.1667em\lower.7ex\hbox{E}\kern-.125emX}}

\def\fig{Figure~}

\definecolor{dkgreen}{rgb}{0,0.6,0}
\definecolor{gray}{rgb}{0.5,0.5,0.5}
\definecolor{mauve}{rgb}{0.58,0,0.82}
\definecolor{mygray}{gray}{0.9}

\lstdefinestyle{top}{
  float=t,
  floatplacement=t,
  abovecaptionskip=2pt
}

\newcommand*\circled[1]{\tikz[baseline=(char.base)]{
            \node[shape=circle,fill, inner sep=0pt, minimum width=0.3cm] (char) {\textcolor{white}{#1}};}}

\definecolor{capcolor}{RGB}{51,51,204}

\newtoggle{todo}
\toggletrue{todo}

\newcommand{\update}[1]{\textcolor{black}{#1}}

\newtoggle{comment}
\togglefalse{comment} 

\newcommand{\softmax}{\mathrm{softmax}}

\newcommand{\defeq}{:=}

\newcommand{\vQ}{\mathbf{Q}}
\newcommand{\vK}{\mathbf{K}}
\newcommand{\vV}{\mathbf{V}}

\newcommand{\vS}{\mathbf{S}}

\newcommand{\vP}{\mathbf{P}}

\newcommand{\vO}{\mathbf{O}}

\SetAlgoSkip{smallskip}

\AtBeginDocument{%
  \providecommand\BibTeX{{%
    Bib\TeX}}}

\setcopyright{none} 
\copyrightyear{2025}
\acmYear{2025}
\acmDOI{XXXXXXX.XXXXXXX}

\acmConference[MICRO 2025]{The 58th IEEE/ACM International Symposium on Microarchitecture}{October 18--22, 2025}{Seoul, Korea}

\acmISBN{978-X-XXXX-XXXX-X/XX/XX}

\begin{document}

\setcounter{page}{1} 

\title{PAM: Processing Across Memory Hierarchy for Efficient KV-centric LLM Serving System}

\author{Lian Liu}
\authornote{Both authors contributed equally to this research.}
\author{Shixin Zhao}
\authornotemark[1]
\affiliation{%
    \institution{Institute of Computing Technology, CAS, University of Chinese Academy of Sciences}
  \city{Beijing}
  \country{China}
  \postcode{100190}
}
\email{liulian211,zhaoshixin18@mails.ucas.ac.cn}

\author{Yutian Zhou}
\affiliation{%
  \institution{Institute of Computing Technology, Chinese Academy of Sciences, University of Chinese Academy of Sciences}
  \city{Beijing}
  \country{China}
}
\email{zhouyutian23s@ict.ac.cn}

\author{Yintao He}
\affiliation{%
  \institution{Institute of Computing Technology, Chinese Academy of Sciences, University of Chinese Academy of Sciences}
  \city{Beijing}
  \country{China}}
\email{heyintao19z@ict.ac.cn}

\author{Mengdi Wang}
\affiliation{%
  \institution{State Key Lab of Processors, Institute of Computing Technology,  Chinese Academy of Sciences}
  \city{Beijing}
  \country{China}}
\email{wangmengdi@ict.ac.cn}

\author{Yinhe Han}
\affiliation{%
  \institution{State Key Lab of Processors, Institute of Computing Technology,  Chinese Academy of Sciences}
  \city{Beijing}
  \country{China}}
\email{yinhes@ict.ac.cn}

\author{Ying Wang}
\authornote{Corresponding author.}
\affiliation{%
  \institution{State Key Lab of Processors, Institute of Computing Technology,  Chinese Academy of Sciences}
  \city{Beijing}
  \country{China}}
\email{wangying2009@ict.ac.cn}



\begin{abstract}

The widespread adoption of Large Language Models (LLMs) has exponentially increased the demand for efficient serving systems. With growing requests and context lengths, key-value (KV)-related operations, including attention computation and KV cache storage, have emerged as critical bottlenecks. They require massive memory bandwidth and capacity. Unfortunately, existing LLM serving systems, optimized for compute-bound workloads, fail to handle these memory-intensive operations effectively. Even with Processing-In-Memory (PIM) technology, current single-level memory designs cannot simultaneously satisfy the bandwidth and capacity requirements, particularly in cost-sensitive cloud environments.

To address these challenges, we propose Processing Across Memory (PAM), a KV-centric LLM serving system that coordinates heterogeneous PIM-enabled memory devices within a hierarchical architecture. PAM introduces a novel computing paradigm to balance high memory bandwidth with scalable capacity. First, PAM exploits the inherent context locality in KV access patterns to intelligently distribute KV tokens across the memory hierarchy with near-data computing capability. Second, to further exploit context locality, it introduces the PAMattention algorithm, enabling fine-grained parallel attention computation across heterogeneous PIM devices. Finally, PAM incorporates an intra-device KV mapping, inter-device KV migration interface, and an inter-device online KV scheduling algorithm to dynamically balance computational workloads and reduce data transfer overhead.
PAM adopts a full-stack design that orchestrates diverse PIM devices as a unified, hierarchical system to meet both bandwidth and capacity demands. 
By addressing both bandwidth and capacity demands simultaneously, PAM significantly enhances the efficiency and scalability of LLM serving systems, paving the way for cost-effective, high-performance solutions in the era of large-scale AI.

\end{abstract}

\maketitle

\section{Introduction}

Large Language Models (LLMs) \cite{achiam2023gpt, touvron2023llama, touvron2023llama2} have emerged as a transformative force in generative AI, revolutionizing applications such as chatbots\cite{chiang2023vicuna, chiang2024chatbot, guo2025deepseek}, search engines~\cite{kabakucs2024battle, xiong2024search}, and automated customer service~\cite{roziere2023code, peng2023study}. The rapid adoption of LLM-based solutions has led to an exponential increase in demand for efficient serving systems, with modern deployments handling millions of requests daily~\cite{requests2024}. This surge in demand has driven the development of various LLM serving systems~\cite{kwon2023efficient, yu2022orca, zhong2024distserve, patel2024splitwise, liu2024cachegen, liu2024comet, li2024specpim, qin2025mooncake, agrawal2024taming}, which aim to optimize serving performance in increasingly resource-intensive scenarios.

\begin{figure*}
    \centering
    \includegraphics[width=\linewidth]{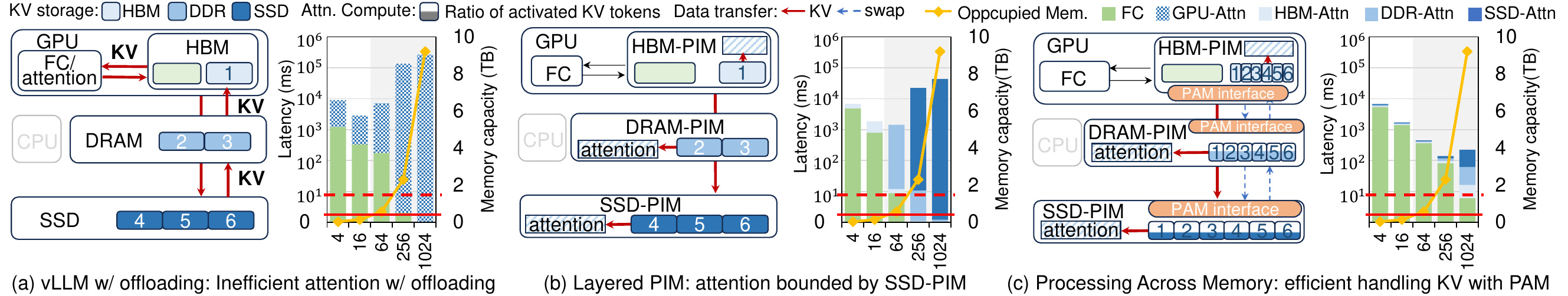}
    \vspace{-0.3cm}
    \caption{Illustration of (a) vLLM with offloading; (b) Layered PIM; and (c) PAM. System illustration is on the left side of each subfigure and the performance breakdown is on the right side of each subfigure. The x-axis of the histogram is batch size. The red solid line indicates the maximum capacity of HBM and the red dotted line indicates the maximum DDR capacity.}
    \label{fig:intro}
\vspace{-0.3cm}
\end{figure*}

Despite these advancements, existing solutions are constrained by two major limitations. First, many systems rely on large-scale, high-performance hardware infrastructures, such as tightly interconnected GPU clusters and expensive High Bandwidth Memory (HBM), to improve processing capability. While effective, these approaches often come with exorbitant costs and energy consumption, making them impractical for cost-sensitive environments. Second, these systems are predominantly compute-centric, prioritizing the optimization of compute-bound operations. However, key-value (KV)-related operations, including attention computation and KV cache storage, constitute the primary bottleneck in LLM serving systems. Attention computation, which generally accounts for over 80\% of runtime, is inherently bandwidth-bound as it requires loading all previously generated KV (referred to as KV-tokens in this paper) to generate each new token~\cite{vaswani2017attention}. KV cache, storing precomputed key-value vectors to avoid redundant computations, can occupy over 95\% of the total memory capacity~\cite{kwon2023efficient, hong2024flashdecoding++, li2024specpim}. Together, these operations demand both high memory bandwidth and vast memory capacity, creating a fundamental mismatch with compute-optimized designs.

Recent studies have explored various hardware~\cite{park2024attacc, zhou2022transpim, wu2024pim, park2024lpddr, pan2024instinfer} and software~\cite{aminabadi2022deepspeed, alizadeh2023llm, mlc-llm, jiang2024neo} techniques to alleviate these bottlenecks. Among these, Processing-In-Memory (PIM) architectures\footnote{For simplicity, DRAM-based and Flash-based PIM/Near Data Processing are collectively referred to as PIM in this paper.}~\cite{wu2024pim, seo2024ianus, kim2023samsung, lee20221ynm, nayak2024fusemax, kwon2022system} have gained prominence. These architectures integrate computational units directly into memory devices, providing increased bandwidth. Systems such as AttAcc!~\cite{park2024attacc}, which leverage HBM-PIM~\cite{lee2021hardware}, achieve up to 9$\times$ higher aggregated memory bandwidth than NVIDIA DGX A100~\cite{dgx-a100}, significantly improving attention computations performance. However, the high cost of large-scale HBM-PIM deployments renders them impractical for scaling KV cache capacity to handle millions of requests. Alternatively, as shown in \fig \ref{fig:intro}(a), offloading-based approaches~\cite{lee2024infinigen, sheng2023flexgen, song2023powerinfer, xue2024powerinfer, hwang2024pre} transfer KV data to inexpensive host memory (e.g., DDR) or non-volatile memory (e.g., SSDs) to expand memory capacity. While cost-effective, these systems suffer from limited bandwidth, as SSDs typically offer less than 5 GB/s compared to the 2 TB/s bandwidth of GPUs with HBM. While these approaches have shown promise, they fail to simultaneously address the dual demands for bandwidth and capacity.


Instead of relying on isolated memory levels or compute-centric optimizations, we argue for a hierarchically coordinated approach that leverages the strengths of heterogeneous memory devices. 
A straightforward solution is to enable in-situ processing in the current hierarchical memory system, i.e. layered processing-in-memory (L-PIM) system as we defined. As shown in \fig\ref{fig:intro}(b), L-PIM distributes KV tokens across HBM, DDR, and SSD, providing large capacity and enhanced bandwidth~\cite{lockerman2020livia, schwedock2022tako, wang2023affinity, fujiki2022multi, xiao2018prometheus}. However, simply putting PIM devices together shows that a new bottleneck emerges in attention computation, e.g. SSDs (SSD-Attn), as memory usage increases. This bottleneck arises from a mismatch between the memory access intensity and the provided bandwidth of different memory devices in the L-PIM system. 
Specifically, since all KV tokens participate in attention computation, the lowest-level memory, such as SSDs, which store the majority of KV tokens, endures the highest frequency of data accesses. However, their limited bandwidth poses significant challenges for efficient data retrieval, leading to substantial performance degradation. For example, SSD-PIM solutions~\cite{jang2024smart, li2021glist, mansouri2022genstore, li2023ecssd} provide a bandwidth of less than 100 GB/s—merely 5\% of the bandwidth offered by HBM-PIM, highlighting the stark disparity in performance capabilities across the memory hierarchy.

This limitation underscores the need for a fundamentally new approach that goes beyond simply combining heterogeneous PIM devices. Processing Across Memory (PAM) addresses this challenge by coordinating KV-related operations across the memory hierarchy in a context-aware, algorithmically optimized manner, as illustrated in \fig \ref{fig:intro}(c). PAM adopts a full-stack design to align algorithms, architectures, and system management with the unique demands of KV-related operations. Leveraging multi-step strategies, PAM overcomes the inefficiencies that limit hierarchical PIM systems.

First, PAM identifies and exploits the \textbf{context locality} inherent in KV caches. Recent studies reveal that attention mechanisms are inherently sparse~\cite{deng2024attention, lee2024infinigen, ribar2023sparq, yang2024post, zhang2024pqcache}, requiring only a subset of KV tokens for subsequent token generation. By analyzing this sparsity, PAM uncovers a strong context locality pattern: KV tokens accessed in recent decoding steps are more likely to be reused in adjacent token generations (\textsection \ref{sec:context-locality}). This insight enables PAM to efficiently distribute KV tokens across the memory hierarchy according to their access patterns. For example, frequently accessed tokens are allocated to HBM, optimizing data placement to leverage the memory hierarchy effectively.

Second, the context locality necessities partitioning KV tokens onto heterogeneous PIM devices, which cannot be achieved through existing request-level solutions~\cite{aminabadi2022deepspeed, yu2022orca, heo2024neupims, park2024attacc}.
PAM achieves fine-grained distributed attention orchestration by introducing the PAMattention algorithm.  
Specifically, PAMattention achieves token-wise parallelization across hierarchical PIM devices by dividing each attention computation into local and global stages  (\textsection \ref{sec:pam-attention}), enabling each PIM device to process its allocated KV tokens locally while contributing to the overall computation in parallel. 
In addition, the distributed attention model introduces additional reduction requirements, which PAM addresses with hierarchical reduction units that efficiently handle intra- and inter-device reductions with minimal overhead. This architecture ensures seamless collaboration across heterogeneous PIM devices, delivering scalable performance for memory-intensive attention operations.

Third, PAM implements a KV-centric management strategy to harness the potential of context locality and balance workloads across the memory hierarchy (\textsection \ref{sec:kv-management}). The management mechanism comprises three components, including intra-device KV mapping, PAM interface, and inter-device KV scheduling. First, the intra-device KV mapping employs a PIM-aware address mapping mechanism to balance workloads and maximize processing capabilities within each PIM device. Second, since the data addressing and layouting mechanism for each PIM device is quite different, we introduce a novel PAM interface to facilitate efficient data layout transformations during KV token transfers across memory layers, minimizing transfer overhead. Finally, based on real-time changing patterns of KV accessing, we propose a novel inter-device KV scheduling strategy to achieve dynamic load balancing across heterogeneous PIM devices with minimal data transfer overhead. 

Figure 1(a)-(c) visualizes the offloading system designs of L-PIM and PAM, highlighting how PAM effectively addresses key bottlenecks in LLM serving. Performance breakdowns illustrate that PAM delivers significant improvements in meeting modern LLM deployment demands. In general, we make the following contributions.
\begin{itemize}[leftmargin=*, nosep]
  \item We present PAM, a scalable LLM serving system that unifies heterogeneous PIM tiers, including HBM-PIM, DRAM-PIM, and SSD-PIM, into a hierarchical memory-compute architecture. Instead of treating each PIM tier in isolation, PAM adopts a top-down co-design approach that orchestrates execution across all tiers. By aligning KV-related operations with tiered architecture, PAM balances bandwidth and capacity demands while maintaining cost efficiency.
  
  \item To reduce inter-tier data movement and exploit hierarchical PIM compute capabilities, we propose PAMattention—a hardware-software co-designed attention algorithm optimized for tiered memory. PAMattention enables fine-grained, token-level parallelism by performing in-situ softmax and embedding dedicated reduction units in each memory tier, reducing communication to achieve efficient attention across heterogeneous PIM tiers.

  \item PAM exploits intrinsic context locality to coordinate cross-tier data placement and movement through a suite of KV management mechanisms. Specifically, PAM utilizes intra-device KV mapping to maximize processing parallelism in each PIM module, while providing a unified inter-device PAM interface and scheduling strategy to minimize the migration of KV tokens across the memory hierarchy. 

  \item Compared to the industry-standard DGX-H100 based vLLM~\cite{kwon2023efficient} serving system, PAM achieves an average \textbf{12.88}$\times$ and \textbf{26.41}$\times$ performance improvement for conversation and long-context processing, respectively, showcasing its significant advantages in throughput and cost-effectiveness for LLM serving.

\end{itemize}
\section{Background}
\begin{figure}
    \centering
    \includegraphics[width=0.98\linewidth]{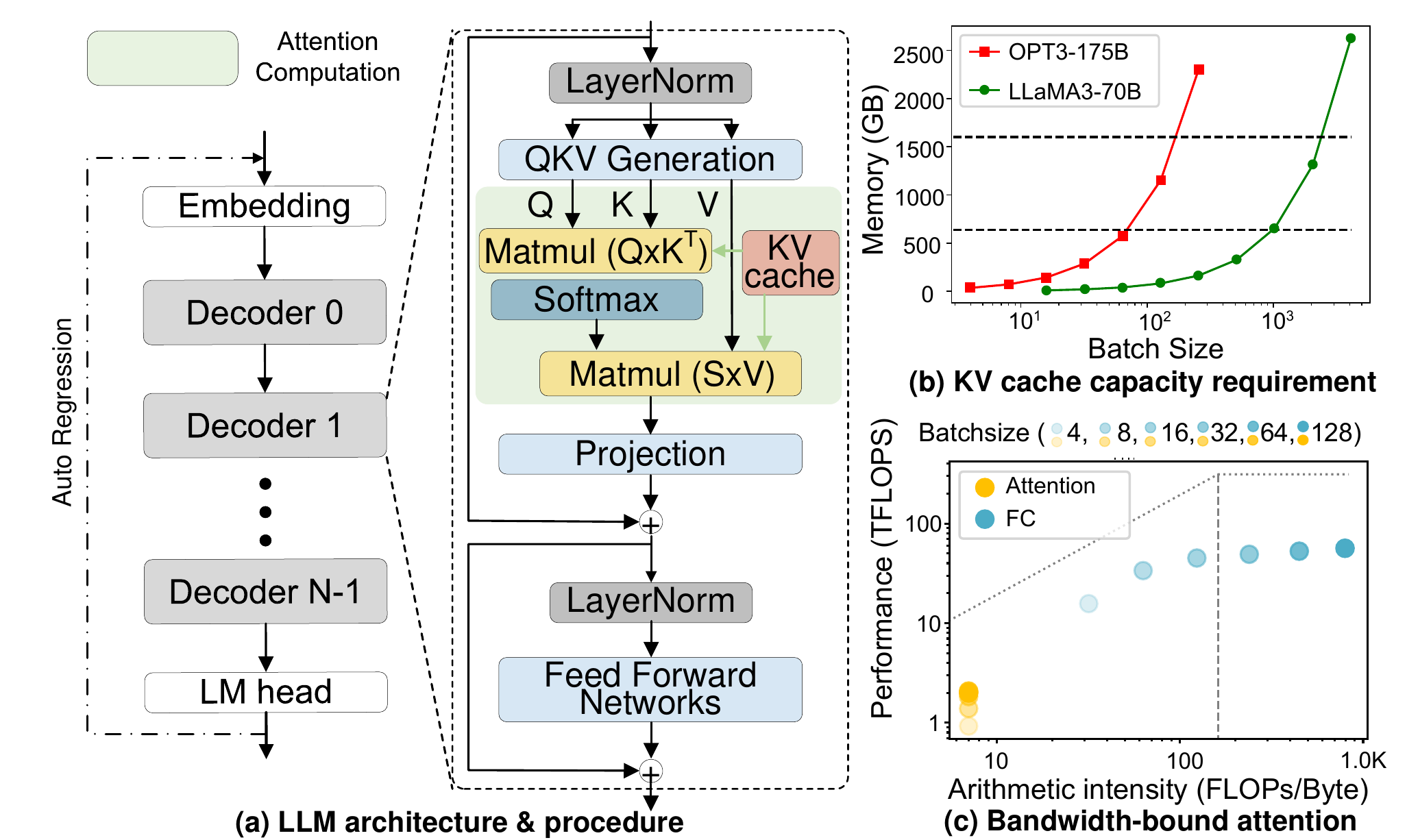}
    \vspace{-0.3cm}
    \caption{(a) LLM architecture. (b) \update{The requirements of LLM serving.} (c) Roofline model of attention computation. }
    \label{fig:llm-illustration}
\vspace{-0.3cm}
\end{figure}

\subsection{LLM Architecture \& Inference}

\subsubsection{LLM Architecture} 
As shown in \fig\ref{fig:llm-illustration}(a), generative LLMs are constructed with multi-layered transformer blocks~\cite{vaswani2017attention}. There are two major operators in a transformer block: fully connected (FC) in blue and attention in green. These two components dominate the computational and memory overhead during inference, making their optimization critical for efficient performance.

\subsubsection{LLM Inference Procedure}
LLM inference operates in two phases: prefill and decoding. In the prefill phase, the model processes  $m$ user-input tokens in parallel to generate the first new token $x_{m+1}$, leveraging the parallelism of modern hardware such as NVIDIA H100 GPUs~\cite{h100paper}. In the decoding phase, LLM leverages an autoregressive manner, wherein each generated output token serves as the input for generating the next token. Furthermore, to avoid redundant computations, key and value (KV) vectors of previously generated tokens, known as KV cache, are stored to be reused in subsequent tokens. For example, LLM takes the $t-$th token as input and utilizes the KV cache of preceding tokens $1 \sim t$ to generate the $(t+1)-$th token. 

\subsection{LLM Serving Requirements}\label{sec:serving-requirements}
Efficient LLM serving often involves batching multiple user requests to maximize hardware utilization. However, handling batched KV-related operations introduces significant demands for both memory capacity and bandwidth.

\subsubsection{Large Capacity for KV Cache Storage}

The KV cache size grows linearly with batch size, creating significant storage pressure. \update{For instance, processing 256 OPT-175B requests with 2048-token contexts requires approximately 2304 GB of KV cache}, which far exceeds the capacity of platforms like DGX-H100~\cite{dgx-h100}. This storage bottleneck severely limits large-scale deployment feasibility.


\subsubsection{High Bandwidth for Attention Computation}
As illustrated in \fig\ref{fig:llm-illustration}(c), attention operations are inherently memory bandwidth-bound as the arithmetic intensity of attention does not scale with various batch size~\cite{seo2024ianus, park2024attacc}. To make it worse, the loading cost of KV cache grows linearly with the context length since the attention operator requires all pre-generated KV tokens during decoding.

\subsection{Optimizations on KV-related Operations}\label{sec:kv-optimizations}

To address the challenges of KV-related operations, researchers have developed solutions spanning algorithmic, software, and hardware innovations.



\subsubsection{KV Sparsity}

Due to the linguistic distribution properties of natural language~\cite{schank1972conceptual, allen1988natural}, attention is inherently sparse~\cite{deng2024attention}. Researchers have explored various sparsity strategies to optimize KV usage, which can be categorized into two types: eviction-based sparsity and retrieval-based sparsity. Eviction-based KV sparsity strategies~\cite{xiao2024duoattention, shi2024keep, gao2024cost, jiang2024minference, cai2024pyramidkv, zhang2023h2o} store only a subset of important tokens by evaluating token significance during KV cache generation and evicting less critical ones. While significantly reducing KV cache storage overhead, these methods often degrade model accuracy by permanently discarding tokens that may become important in subsequent decoding steps. In contrast, retrieval-based sparsity strategies~\cite{deng2024attention, lee2024infinigen, ribar2023sparq, yang2024post, tang2024quest} store the entire KV cache but selectively load the most relevant tokens during attention computation. However, retrieval-based methods do not reduce storage footprint, as they still cache the full KV set.


\subsubsection{PIM-based Acceleration}
Processing-In-Memory (PIM) architectures address the bandwidth demands of attention computations by integrating computational capabilities directly into memory devices~\cite{lee20221ynm, zhou2022transpim, kim2023samsung, kim2023optimstore, liang2019cognitive}. Two common design approaches are: near-bank/plane and on-controller designs.
In near-bank PIM designs, processing units are placed close to the memory banks. This design enables operations to be performed directly using data read from the row buffers, fully exploiting the internal bandwidth of multiple memory banks to process bandwidth-bound attention operators. For example, AttAcc!~\cite{park2024attacc} leverages the high aggregate memory bandwidth of HBM-PIM~\cite{lee2021hardware} for efficient attention computations. However, near-bank PIM design cannot achieve efficient data communication across different banks~\cite{gomez2021benchmarking, noh2024pid, tian2024ndpbridge}.

On-controller designs~\cite{zhou2023dimm, liang2019cognitive} centralize computation within the memory controller, enabling operations such as reductions without data movement. While this improves efficiency in certain cases, it provides lower bandwidth than near-bank PIM approaches.




\subsubsection{Memory Management}
To address the significant storage requirements of KV caches, memory management strategies have been developed to efficiently utilize fragmented resources across heterogeneous devices~\cite{prabhu2024vattention, xu2024vtensor, sheng2023flexgen, maurya2024breaking, he2024fastdecode, chen2024efficient, niu2024smartmem}. For example, vLLM introduces a paged attention mechanism that maps KV caches into multiple blocks within GPU memory, reducing fragmentation~\cite{kwon2023efficient}. However, GPU memory constraints still limit scalability. 
To address this issue, offloading-based strategies provide additional capacity by transferring portions of the KV cache to CPU memory or SSDs~\cite{sheng2023flexgen, lee2024infinigen, jain2022hugging}. For example, DeepSpeed-Inference~\cite{aminabadi2022deepspeed} offloads KV cache to host memory to expand available capacity. This approach effectively increases the available serving capacity. However, such systems must transfer KV caches back to GPU memory for computation, creating bottlenecks due to the limited bandwidth of interconnects like PCIe that bridge heterogeneous memory devices, making it challenging for offloading-based heterogeneous systems to achieve efficient LLM serving.

\section{Motivation \& Challenges}\label{sec:moti_chall}
\subsection{Limitations of Layered PIM}\label{sec:lpim-limitations}

Existing approaches, including offloading-based strategies~\cite{lee2024infinigen, sheng2023flexgen, song2023powerinfer, xue2024powerinfer, hwang2024pre} and PIM architectures within single-level memory device~\cite{wu2024pim, seo2024ianus, pan2024instinfer, yu2024cambricon, zhou2022transpim, he2020newton}, fail to address the dual challenges of massive memory capacity and high bandwidth required for efficient LLM serving. L-PIM systems offers a promising alternative~\cite{dao2022flashattention, alpern1994uniform, wu2021storage, cohen1989storage, he2024fastdecode, aminabadi2022deepspeed, fujiki2022multi}. However, as shown in \fig\ref{fig:intro}(b), L-PIM systems do not significantly improve LLM serving efficiency.

The main limitation arises from the mismatch between KV-related operation patterns and the processing capabilities of L-PIM systems. Attention computation requires loading all pre-generated KV tokens when generating a new token. Due to the large capacity of KV caches, most tokens are offloaded to cost-effective memory like SSDs. This results in an inherent bottleneck: the memory device storing most KV tokens experiences the highest access pressure, yet devices with the largest storage capacity offer the lowest memory bandwidth, far below the requirements of KV-related operations~\cite{jang2024smart, li2021glist, mansouri2022genstore, li2023ecssd}. This inefficiency arises as the traditional memory hierarchy is optimized for exploiting data distribution locality, which is absent in KV operations. KV tokens used in attention computation are treated uniformly, regardless of their relevance or locality. This leads to imbalanced memory access and underutilization of bandwidth across the hierarchy, preventing L-PIM systems from realizing their theoretical performance potential.


\subsection{Context Locality}\label{sec:context-locality}

\begin{figure}
    \centering
    \includegraphics[width=0.9\linewidth]{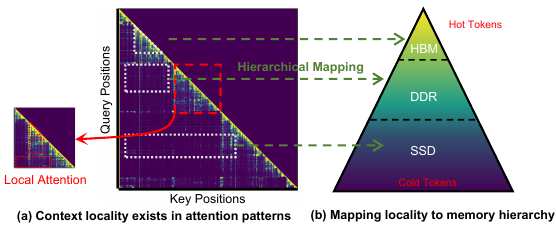}
    \vspace{-0.3cm}
    \caption{\update{Exploiting context locality to achieve hierarchy memory processing.}}
    \label{fig:locality}
\vspace{-0.3cm}
\end{figure}

Fortunately, we observed that existing retrieval-based KV sparsity patterns exhibit strong \textbf{context locality}, where the same KV tokens are frequently accessed within a short period during decoding. For instance, if a KV token $i$ is accessed at decoding step $j$, it is highly likely to be accessed again at step $j+1$. This arises from the semantic properties of natural language, where adjacent tokens are often semantically linked, maintaining tight relationships with specific KV tokens~\cite{zhang2024cam, shazeer2019fast, liu2024minicache}. As illustrated in \fig\ref{fig:locality}(a), we analyzed the attention pattern
of layer-12 in LLaMA2-7B~\cite{touvron2023llama2}. 
\update{The most relevant KV tokens are denoted in yellow, while the KV tokens in dark blue are useless.} The results indicate that a small subset of KV tokens plays a key role in accuracy, while the rest are sporadically activated. Furthermore, these critical KV tokens tend to cluster around the current token, reflecting their local relevance and semantic proximity.

This context locality motivates us to leverage it within the memory hierarchy, to process KV-related operations across the memory hierarchy. As illustrated in \fig\ref{fig:locality}(b), we leverage the activated frequency to guide the mapping of KV tokens to different layers of the PIM hierarchy. It enables collaborative attention computation across PIM devices, optimizing both efficiency and performance.

\subsection{Challenges}\label{sec:pam-challenges}

To realize the proposed Processing-Across-Memory (PAM) paradigm and fully exploit the benefits of context locality, the following key challenges must be addressed:

\subsubsection{C1: Inefficient Attention Computation within Heterogeneous PIM Devices\label{sec:c1}}

As shown in \fig\ref{fig:llm-illustration}(a), attention computation involves three main kernels: two matrix multiplications (matmul) interleaved with a softmax operation. Unlike matmul, which benefits from efficient tiling~\cite{shen2021efficient, zheng2023chimera}, the all-to-all nature of softmax complicates distributed execution. While KV tokens can be mapped to heterogeneous devices based on activation frequencies, and tiling can optimize the matmul $Q \times K^{T}$ within each device, softmax requires gathering all outputs onto a single device. The results must then be redistributed for the matmul $S \times V$, incurring significant overhead due to limited inter-device bandwidth. Minimizing data movement while maintaining computational efficiency is critical.

\subsubsection{C2: Dynamic Distribution of Activated KV Tokens\label{sec:c2}}

Although context locality exists, the distribution of activated KV tokens evolves dynamically during decoding. As shown in \fig\ref{fig:locality}(a), their importance shifts over time, influenced by the changing context of token generation. A static KV mapping results in imbalanced memory access, with some devices overloaded while others remain underutilized. To mitigate this, an online scheduling mechanism is needed to predict KV token importance and dynamically redistribute them across the memory hierarchy.

\section{\update{PAM: System Overview}}\label{sec:system-overview}
\begin{figure*}
    \centering
    \includegraphics[width=0.98\linewidth]{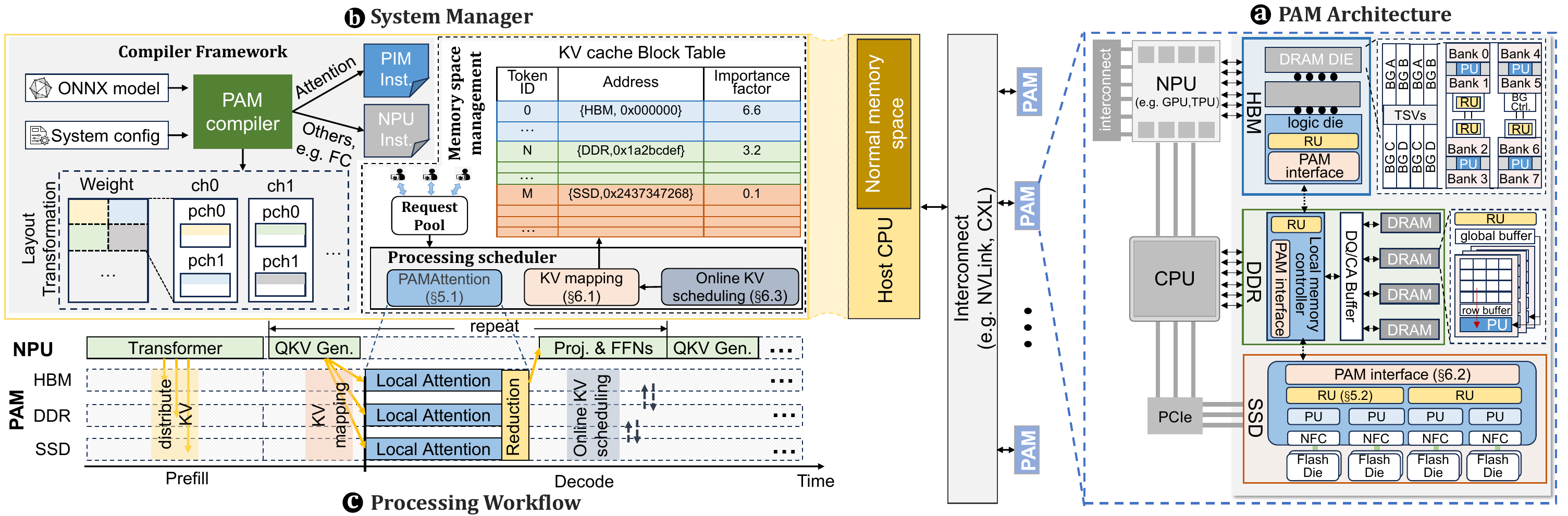}
    \vspace{-0.3cm}
    \caption{\update{Overview of the proposed KV-centric LLM serving system, PAM.}} 
    \label{fig:pam-overview}
\vspace{-0.3cm}
\end{figure*}


\update{We present PAM, an efficient KV-centric LLM serving system that introduces the novel Processing-Across-Memory paradigm to address the challenges mentioned above. Figure \ref{fig:pam-overview} illustrates the overview of PAM system, which comprises three core components: a tiered memory hierarchy with PIM capabilities, a centralized system manager, and a heterogeneous processing coordination framework.}

\subsection{\update{System Architecture}}\label{sec:pam-architecture}

\update{As presented in Figure \ref{fig:pam-overview}\circled{a}, PAM system builds upon modern LLM serving platforms such as DGX-H100~\cite{dgx-h100}, employing high-bandwidth interconnects (NVLink/CXL) between multiple PAM instances. Each instance integrates CPU resources with NPU (e.g., GPUs~\cite{a100paper, h100paper}, TPUs~\cite{jouppi2023tpu, jouppi2017datacenter}) through a multi-tier PIM-enabled memory hierarchy spanning HBM, DDR, and SSD. NPUs connect to HBM via 2.5D packaging, while CPUs interface with DDR and SSD through standard memory buses and PCIe/NVMe, respectively.}

\update{PAM innovatively introduces a tiered PIM design combining HBM-PIM, DDR-PIM, and SSD-PIM to accelerate memory-intensive KV-related operations. HBM-PIM and DDR-PIM adopt near-bank processing units (PUs) to exploit DRAM bandwidth, each tier equipped with dedicated reduction units (RUs) for efficient aggregation of variable-length attention (details in~\textsection\ref{sec:heterpim-architecture}). SSD-PIM employs controller-level PU/RU designs to parallelize attention across flash dies. To support cross-tier access across heterogeneous layouts, each PIM device integrates a PAM interface (see~\textsection\ref{sec:pam-interface}) to facilitate efficient data transformation and transfer.}

\update{While Figure \ref{fig:pam-overview} depicts a single-node configuration, PAM seamlessly scales to multi-node deployments. Following strategies in existing large-scale serving systems~\cite{wu2024loongserve, qin2025mooncake}, the system achieves this through RDMA networking (8$\times$400 Gbps interconnects). It also supports hybrid tensor/pipeline parallelism for efficient serving. \textsection\ref{sec:exp-scalability} shows near-linear scaling across instances and nodes.}

\subsection{\update{System Manager}}
\update{As illustrated in Figure \ref{fig:pam-overview}\circled{b}, PAM incorporates a dedicated System Manager that orchestrates: (1) a compiler framework, (2) Memory and storage space management, and (3) Cross-tier execution scheduling for efficient LLM serving. This management stack is built upon the application level without OS modification.}

\subsubsection{\update{Compiler Framework. }}
\update{
The PAM compiler serves as the system's frontend, ingesting LLM model specifications (in ONNX format~\cite{onnx-model}) alongside hardware configuration to produce an executable inference workflow. It performs operator graph parsing, tile sizing, and instruction scheduling to optimize resource usage. Attention operators are compiled into tiered-PIM instructions, while non-memory-bound operators are mapped to NPU kernels.
In addition, the compiler performs offline weight layout transformation, to ensure high-throughput access to weights stored in PIM devices and that weight fetching aligns with NPU execution needs.} 


\subsubsection{\update{Memory/Storage Space Management. }}
\update{Aligned with commercial PIM practices~\cite{lee2021hardware, devaux2019true, mailthody2019deepstore}, PAM manages all PIM-enabled memories (HBM-/DDR-/SSD-PIM) through physical address control, bypassing OS-level abstractions like virtual memory, page faults, or caching. PAM organizes three main types of data: KV cache, model weights, and intermediate activations. For KV cache, PAM adopts PagedAttention~\cite{kwon2023efficient}, using a block table to record the physical locations of KV tokens (see~\textsection\ref{sec:kv-allocation}). To preserve full compute locality within PIM devices, PAM disables address hashing, similar to techniques used in commercial implementations~\cite{devaux2019true, lee20221ynm}. While this limits CPU/NPU direct access efficiency, separate data placement strategies are used to mitigate the impact. Following Attacc!~\cite{park2024attacc}, intermediate activations are mapped directly to memory buffers across tiers, avoiding interference from address hashing. For large-size weights, PAM applies the compile-time layout transformation strategy as in UM-PIM~\cite{zhao2024pim} to ensure high-bandwidth utility. The host CPU manages both control flow and the KV block table. Evaluation shows that this management requires less than 4GB per node, so PAM reserves an 8GB region on host memory—comparable to practices in UPMEM and HBM-PIM deployments.}

\subsubsection{\update{Processing Scheduler. }}
\update{PAM maintains a request pool for incoming queries and schedules execution at runtime. The scheduler prioritizes prefill stages, following vLLM’s policy~\cite{vllm-project}, and manages PAMattention computing and KV token placement across tiered-PIMs to optimize locality and computation (details in~ \textsection\ref{sec:heterattention} and \textsection\ref{sec:kv-scheduling}). Each KV token is tagged with an importance factor, stored alongside the block table to guide runtime scheduling.}

\subsection{\update{Processing Workflow}}\label{sec:workflow}


\update{PAM orchestrates NPUs for compute-intensive tasks and delegates memory-bound operations to tiered PIM modules for efficient end-to-end inference. As shown in Figure~\ref{fig:pam-overview}\circled{c}, execution proceeds in two phases. During prefill, NPUs run all operators~\cite{patel2024splitwise, zhong2024distserve} while distributing KV cache across memory tiers. In decoding, the host offloads QKV generation to NPUs, then transfers the resulting Q, K, and V vectors to tiered PIM modules. The host triggers local attention computations on each PIM tier, with partial results aggregated in HBM-PIM, where reduction units are used to complete the full attention computation (see~\textsection\ref{sec:pam-attention}). The attention output is returned to NPUs for output projection and FFN layers. Meanwhile, the scheduler dynamically adjusts token placement across tiers.}

\section{PAMattention: Attention across Memory}\label{sec:pam-attention}
The challenge outlined in \textsection \ref{sec:c1} demonstrates that attention computation across memory tiers incurs substantial data movement overhead despite KV sparsity and context locality. Our PAMattention solution introduces in situ \textit{online softmax} for fine-grained token-wise parallelism within each PIM device, dramatically reducing inter-tier transfers. This section details both the algorithm (\textsection \ref{sec:heterattention}) and its associated architectural implementation (\textsection \ref{sec:heterpim-architecture}).

\subsection{PAMattention Algorithm}\label{sec:heterattention}

\subsubsection{Tiling for Softmax}

While kernel fusion~\cite{zheng2023chimera, cai2022optimus, niu2021dnnfusion, zheng2023einnet} reduces data movement, attention computations resist fusion due to softmax's all-to-all dependencies. To overcome this, we adopt an equivalent transformation method, enabling softmax tiling~\cite{shen2021efficient, dao2022flashattention, flashdecoding} with online scaling. Specifically, the softmax computation for a vector $x \in \mathbb{R}^{1\times d}$ is defined as:
\begin{align}
  & m(x) \defeq \max_i\ \ x_i, \quad
  f(x) \defeq \begin{bmatrix} e^{x_1 - m(x)}, \hdots, e^{x_{d} - m(x)} \end{bmatrix}, \\
  & \ell(x) \defeq \textstyle\sum_i f(x)_i, \quad
  \softmax(x) \defeq \frac{f(x)}{\ell(x)}.
\label{eq:normal-softmax}
\end{align}

By tiling the input vector into sub-vectors $\scriptsize x = \begin{bmatrix} x^{(1)}, x^{(2)} \end{bmatrix} \in \mathbb{R}^{1 \times d}$, the computation can be expressed:
\begin{align}
  &m(x) = m(\begin{bmatrix} x^{(1)}, x^{(2)} \end{bmatrix}) = \max(m(x^{(1)}), m(x^{(2)})), \\
  &f(x) = \begin{bmatrix} e^{m(x^{(1)}) - m(x)} f(x^{(1)}), e^{m(x^{(2)}) - m(x)} f(x^{(2)}) \end{bmatrix}, \\
  &\ell(x) = e^{m(x^{(1)}) - m(x)}\ell(x^{(1)}) + e^{m(x^{(2)}) - m(x)} \ell(x^{(2)}), \\
  &\softmax(x) = \frac{f(x)}{\ell(x)}.
\label{eq:online-softmax}
\end{align}

This decomposition allows for the parallel computation of independent tiles in different PIM devices by maintaining intermediate statistics ($m(x), \ell(x)$). This process, termed \textbf{online softmax}, forms the foundation for our PAMattention algorithm.

\begin{figure}
    \centering
    \includegraphics[width=0.98\linewidth]{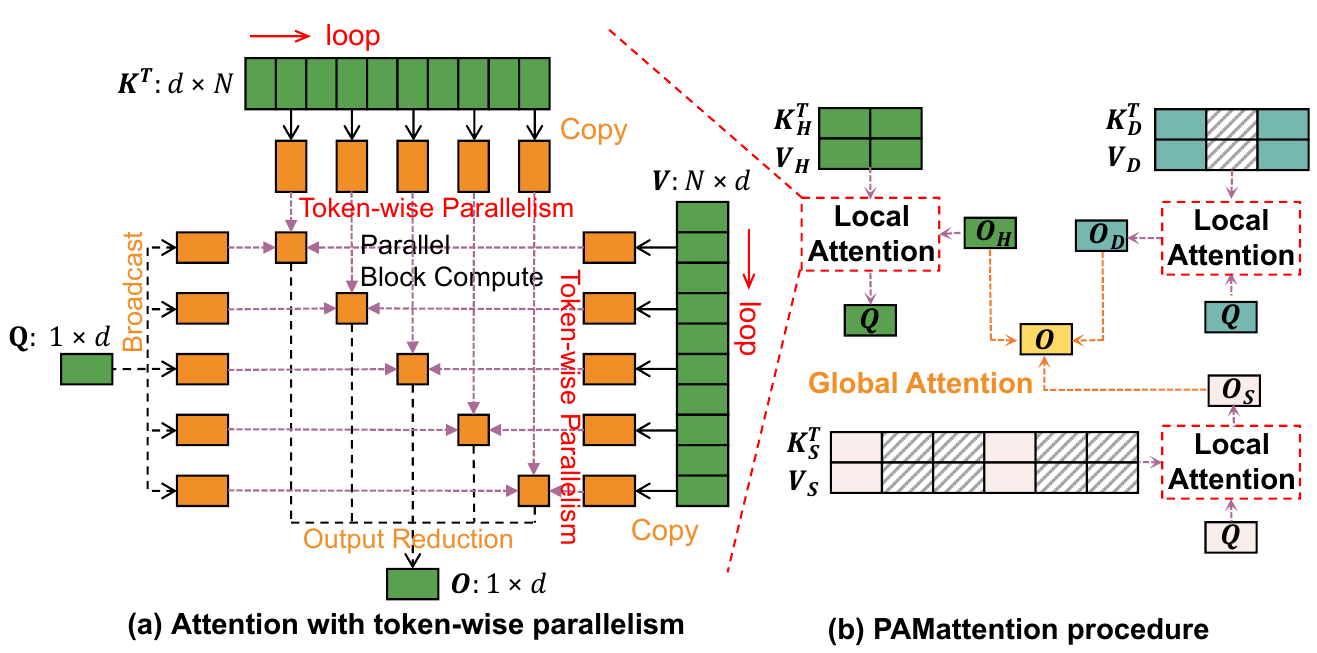}
    \vspace{-0.3cm}
    \caption{Illustration of PAMattention processing.}
    \label{fig:pam-attention}
\vspace{-0.3cm}
\end{figure}

\IncMargin{1em}
\begin{algorithm}[th]
\small
\newcommand{\mycommentstyle}[1]{{\normalsize #1}}
\SetKwComment{Comment}{\mycommentstyle{\hfill $\rhd$}}{}
\caption{PAMattention Procedure}
\label{alg:pamattn} 
\SetKwFunction{LAttention}{Local\_Attention}
\SetKwFunction{Reduction}{Reduction}
\SetKwInOut{Input}{Input}\SetKwInOut{Output}{Output}

\Input{KV Cache Matrices $\vK, \vV \in \mathbb{R}^{N \times d}$, Vector $\vQ \in \mathbb{R}^{1 \times d}$;} 
\Output{Vector $\vO \in \mathbb{R}^{1 \times d}$;}
 \BlankLine 
  Divide $\vK, \vV$ into three blocks $\vK_H, \vK_D, \vK_S$ and $\vV_H, \vV_D, \vV_S$ according to their storage devices;

  \emph{{\textcolor{magenta}{// Parallel Execution on Hetergeneous PIMs}}}\;
 \textcolor{blue}{Parallel} \For{Each block $\vK_i, \vV_i$}{ 
        $\vO_i, m_i, \ell_i$ = \LAttention{$\vQ, \vK_i, \vV_i$};
 }
  $\vec{\vO} = [\vO_H, \vO_D, \vO_S]$, $\vec{m} = [m_H, m_D, m_S]$, $\vec{\ell} = [\ell_H, \ell_D, \ell_S]$\;
  $\vO, m, \ell$ = \Reduction{$\vec{\vO}, \vec{m}, \vec{\ell}$}\;
  Return the output $\frac{\vO}{\ell}$\;
\

\SetKwProg{Fn}{Function}{:}{}
\Fn{\LAttention{$\vQ, \vK, \vV$}}{
    \textcolor{blue}{Parallel} \For{ $1 \le j \le N_i$}{
        $\enspace \vS^{(j)} = \vQ \vK_j^T \in \mathbb{R}^{1 \times B_j}$; \hfill $\rhd$  \textcolor{magenta}{$\vK_j \in \mathbb{R}^{B_j \times d}$}\;
        $\enspace m^{(j)} = \mathrm{rowmax}(\vS^{(j)})$, 
        $\tilde{\vP}^{(j)} = \exp(\vS^{(j)} - m^{(j)})$, $\ell^{(j)} = \mathrm{row sum}(\tilde{\vP}^{(j)})$, $\vO^{(j)} = \tilde{\vP}^{(j)}\vV_j$;
    }
    $\vO_i, m_i, \ell_i$ = \Reduction{$\vec{\vO}, \vec{m}, \vec{\ell}$}\;
}

\

\Fn{\Reduction{$\vec{\vO}, \vec{m}, \vec{\ell}$}}{
    \emph{{\textcolor{magenta}{// Find the index $t$ occupies the largest $m^{(t)} \in \vec{m}$}}}\;
    $m^{(t)} = \mathrm{max}(\vec{m})$\;
    \For{$\vO^{(j)}, m^{(j)}, \ell^{(j)} \in \vec{\vO}, \vec{m}, \vec{\ell}$}{
        $\ell^{(j)} = e^{m^{(j)}-m^{(t)}} \cdot \ell^{(j)}$\; 
        $\vO = \vO + e^{m^{(j)}-m^{(t)}} \cdot \vO^{(j)}$\;
    }
    $\ell = m^{(t)} + \mathrm{log}(\mathrm{sum}(\vec{\ell}))$\;
    Return $\vO, m^{(t)}, \ell$
}

\end{algorithm}

\subsubsection{Token-wise Parallelism}
Building on online softmax, the KV cache is partitioned into independent tiles to enable parallel processing~\cite{hong2024flashdecoding++, dao2024flashattention2}. As shown in \fig\ref{fig:pam-attention}(a), the intermediate outputs are merged through a lightweight reduction step, enabling fine-grained token-wise parallelism. During this procedure, each tile maintains its local maximum and normalization factors $(m, l)$, synchronized during the reduction phase to ensure global consistency.

\subsubsection{Algorithm Design}


The key idea of the PAMattention algorithm is to partition the attention workload according to the KV cache locations, distributing them among HBM ($\scriptsize K_{H}$, $\scriptsize V_{H}$), DDR ($\scriptsize K_{D}$, $\scriptsize V_{D}$), and SSD ($\scriptsize K_{S}$, $\scriptsize V_{S}$). The entire procedure, as detailed in Algorithm ~\ref{alg:pamattn}, includes three steps, local attention computation, intra-device and inter-device reduction. 
First, local attention computation (lines 9-12) is responsible for processing the KV tokens distributed across different locations (e.g., separate banks within the DRAM die). With our token-wise parallelism design, the KV tokens stored at different locations can be processed in parallel. Alongside the two matmul operations ($Q \times K^T$ and $\scriptsize S \times V$), the local function also computes and stores auxiliary scalars $m$ and $l$ for later reduction. 
Second, intra-device reduction is responsible for merging all outputs in the same PIM device, while inter-device reduction handles all outputs from heterogeneous PIM devices in HBM-PIM. During this reduction procedure (lines 15-22), the global maximum $m^{(t)}$ is first identified. We leverage it to adjust the local output (lines 19-20) and accumulate them in the final result.

Overall, this hierarchical approach minimizes inter-device data transfers, reducing communication overhead while maintaining computational efficiency.

\subsection{\update{Architecture Design for PAMattention}}\label{sec:heterpim-architecture}

As depicted in \fig\ref{fig:heterpim-arch}, we propose a multi-tiered design in each PIM device to support PAMattention. The design incorporates local attention Processing Units (PUs) and Reduction Units (RUs) to optimize the computation efficiency of PAMattention.

\subsubsection{\update{Local Attention Processing Unit (PU)}}

\begin{figure}
    \centering
    \includegraphics[width=0.98\linewidth]{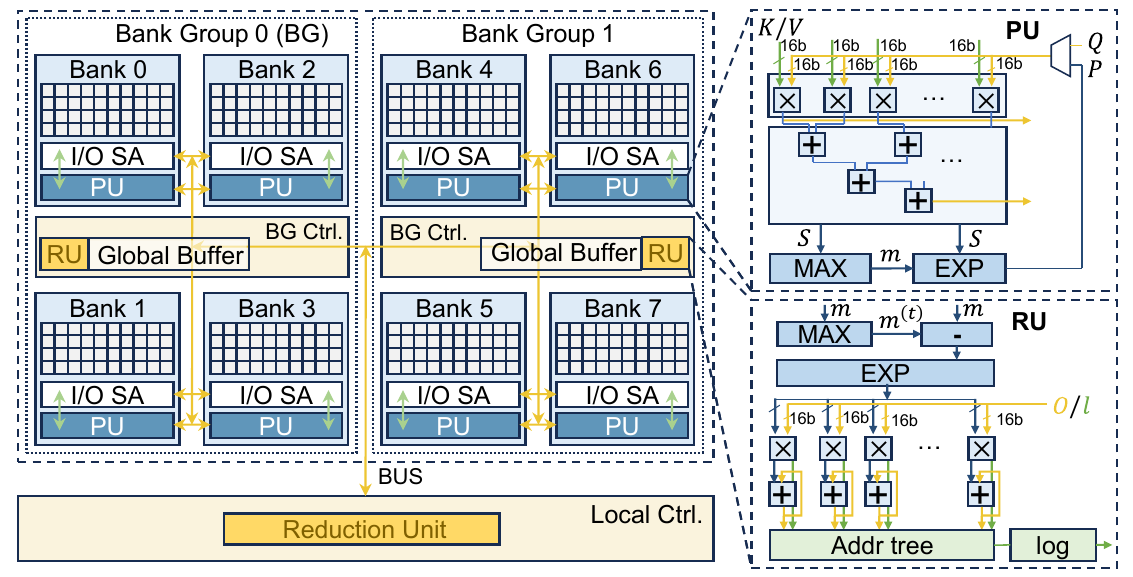}
    \vspace{-0.3cm}
    \caption{Processing unit designed for PAMattention.}
    \label{fig:heterpim-arch}
\vspace{-0.3cm}
\end{figure}

\update{Local Attention Processing Units (PUs) execute the local attention step (line 9) in PAMattention. As shown in the upper-right corner of Figure~\ref{fig:heterpim-arch}, each PU includes a specialized Vector Unit for matrix multiplications ($Q \times K^T$, $S \times V$), built with parallel FP16 multipliers and accumulators. The design is customized per memory tier to balance performance and area constraints:}
\begin{itemize}[leftmargin=*, nosep] 
    \item \update{PAM-HBM PUs are located near-bank within the DRAM die using a DRAM-compatible process\cite{park2024attacc,heo2024neupims,lee2021hardware}. To match the burst access width of HBM, these PUs feature a 256-bit vector unit (16 FP16 multipliers and adders) and are equipped with a 512-byte local buffer for staging input/output results.}
    \item \update{PAM-DDR PUs also adopt near-bank design and utilize a 64-bit vector unit (4 FP16 multipliers and adders) with a 128-byte buffer.} 
    \item \update{PAM-SSD PUs are integrated within the SSD controller. Each controller houses 64 PUs, each equipped with a 256-bit vector unit (16 FP16 multipliers and adders) and a 512-byte buffer.} 
\end{itemize}
\update{Each PU also incorporates dedicated circuits for online softmax evaluation. Specifically, a Row-wise Max Unit identifies the maximum value within each assigned row segment, and a lightweight FP16 exponential unit computes element-wise exponentials.}

\update{During execution, each PU fetches the query vector Q from its local buffer and directly accesses its assigned KV tokens from the local memory bank, enabling efficient in-place computation of vector products and intermediate statistics in online softmax.}

\subsubsection{\update{Reduction Unit (RU)}}

\update{Efficient, scalable reduction is essential for local attention. Without dedicated support, partial outputs must be transferred off-die (e.g., to a logic die), severely underutilizing internal memory bandwidth~\cite{tian2024ndpbridge}. To address this, PAM deploys Reduction Units (RUs) in each memory tier to enable in-place, parallel reductions (line 15), maximizing local bandwidth.}

\update{Each RU includes a 16-FP16 vector-scalar multiplier, comparator, 4 FP16 exponential units~\cite{da2022efficient}, a logarithmic unit, and a 512B scratch buffer for accumulation (Figure~\ref{fig:heterpim-arch}, bottom-right). A lightweight controller handles input staging and pipelined scheduling for asynchronous execution.}
\update{The physical placement of RUs varies on different memory devices. a) PAM-HBM: One RU is deployed per bank group, with an additional 128 RUs integrated into the logic die. 
b) PAM-DDR: One RU is embedded within each DRAM bank group, and 4 additional RUs are placed at the central buffer to aggregate results across chips. c) PAM-SSD: RUs are co-located with the SSD controller. Each controller integrates 8 RUs. }

\update{At runtime, RUs fetch partial outputs from their associated PUs or nearby buffers. It determines the row-wise maximum using the comparator, then computes exponentials and accumulates them in its local buffer. Finally, the logarithmic unit performs normalization based on the accumulated sum. This reduction pipeline is fully overlapped with upstream PU execution, supporting continuous streaming of intermediate results without requiring global synchronization.}
\update{This tiered, distributed RU architecture reduces intra-device transfers by 59\%, cuts reduction latency to <2\% of PAMattention time, and improves aggregate reduction bandwidth by 8–12$\times$ over centralized schemes~\cite{park2024attacc, heo2024neupims}. By aligning reduction with memory locality and supporting pipelined execution, PAM sustains high bandwidth across tiers and scales efficiently.}

\section{PAM: KV-centric Management}\label{sec:kv-management}

The KV-centric management in PAM adopts a token-wise granularity, leveraging token-level parallelism and maximizing context locality benefits. Since KV token placement directly impacts in-memory processing performance, static allocation is suboptimal due to the dynamic nature of activated KV tokens (see \textsection \ref{sec:c2}). To address these challenges, PAM introduces intra-device KV mapping (\textsection \ref{sec:kv-allocation}) to fully utilize the internal bandwidth of each PIM device. Furthermore, we propose a novel PAM interface (\textsection \ref{sec:pam-interface}) to minimize data transfer overhead across different PIM devices. We also employ a greedy-based KV scheduling strategy (\textsection \ref{sec:kv-scheduling}) to balance computing efficiency across the memory hierarchy.

\subsection{Intra-device KV Mapping}\label{sec:kv-allocation}

\begin{figure}
    \centering
    \includegraphics[width=\linewidth]{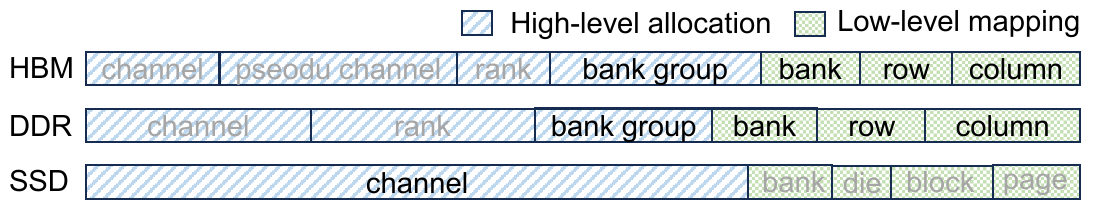}
    \vspace{-0.3cm}
    \caption{\update{The address illustration of each memory hierarchy.}}
    \label{fig:address-mapping}
    \vspace{-0.3cm}
\end{figure}
Efficient KV token mapping within each device is critical for workload balance and reduced processing latency. PAM leverages token-level parallelism by interleaving KV tokens across bank groups (HBM/DDR) or channels (SSDs). To optimize mapping, PAM adopts a two-level strategy aligned with each memory hierarchy, as shown in \fig\ref{fig:address-mapping}. The first level assigns KV tokens to bank groups or SSD channels, while the second refines their placement within the selected group or channel.

For high-level allocation, bank groups (or SSD channels) operate in parallel, making the one with the highest processing latency the bottleneck for intra-device mapping ($T_{intra}$), defined as: $T_{intra} = \max_{bg \in \mathbb{BG}} T_{bg}$,
where $\mathbb{BG}$ represents the set of all bank groups, and $T_{bg}$ denotes the computation latency of a specific bank group.
The number of activated KV tokens influences 
To reduce overall computation latency, PAM balances KV loads across bank groups, as their processing time depends on the number of activated KV tokens, which drive local attention and reduction operations.
Based on context locality, PAM first tracks the activation frequency of KV tokens within each bank group over a specified time window (set to 10 decoding processes). Then, KV tokens are greedily allocated to the bank group with the lowest activation frequency, thus balancing the computation load. This strategy is lightweight: for an 80GB HBM stack with 5 devices and 1,280 bank groups~\cite{park2024attacc}, only 2.8KB is needed for bank group IDs and under 50KB for activation tracking.

After determining the high-level address, PAM refines internal address mapping based on device-specific access patterns. For HBM and DDR with near-bank designs—where all banks can activate simultaneously~\cite{lee2021hardware}—PAM distributes KV tokens evenly across banks to maximize parallel local attention computation. As illustrated in \fig\ref{fig:pam-interface}(a)(b), banks within a group activate the same rows concurrently, so PAM aligns KV tokens to identical rows and columns across banks, enabling efficient parallelism~\cite{wu2024pim, park2024attacc}. Since PAM adopts an on-controller computing design~\cite{wilkeningrecssd} for SSDs, we can easily follow established interleaved storage strategies~\cite{kim2002space, lee2006fast} to boost processing efficiency.

\subsection{Inter-device PAM Interface}\label{sec:pam-interface}

\begin{figure}[t]
    \centering
    \includegraphics[width=0.98\linewidth]{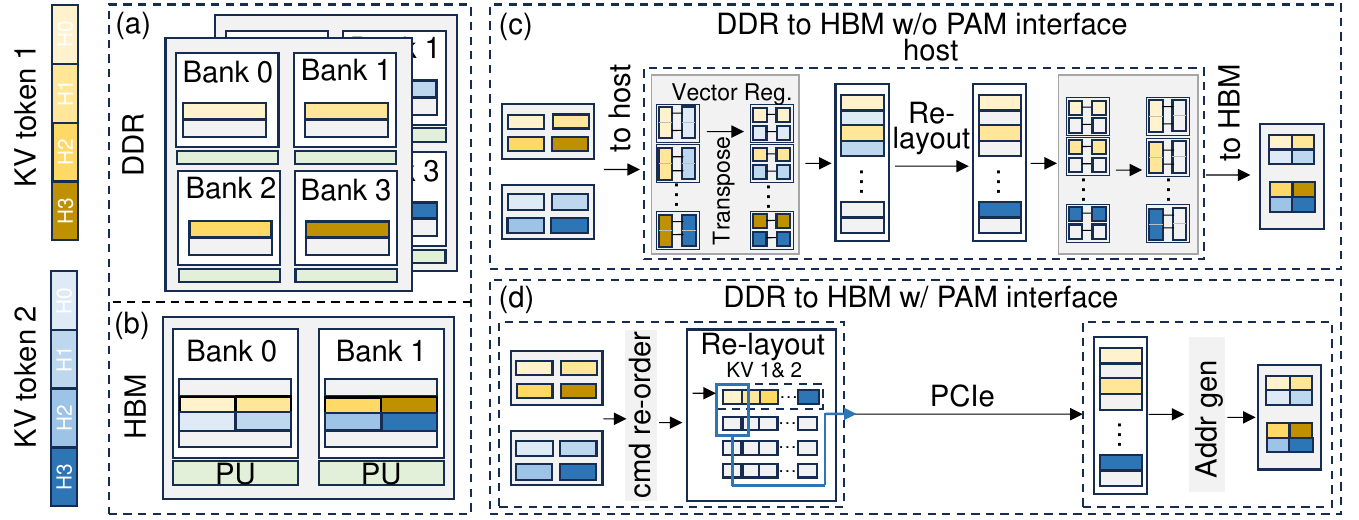}
    \vspace{-0.3cm}
    \caption{(a)(b) KV token mapping across the banks within bank group; (c) Data layout transformation using host CPU causes frequent CPU-PIM interaction; (d) PAM interface enhances efficient data transfer.}
    \label{fig:pam-interface}
\vspace{-0.3cm}
\end{figure}

Efficient intra-PIM computation requires device-specific KV cache layouts, necessitating frequent layout transformations—especially during cross-tier KV transfers. As shown in Figure~\ref{fig:pam-interface}(c), performing these transformations on the host CPU incurs high overhead from CPU-PIM communication and processing latency~\cite{noh2024pid}. \update{This is further exacerbated by PAM’s disabled address hashing, which reduces CPU-side bandwidth to below 10\% of optimized levels. Moreover, layout mismatches across devices force costly data exchanges: data must be read into CPU memory, reformatted, and written back—introducing >500ms latency for MB-scale transfers, which is prohibitive for LLM inference.}

To this end, PAM introduces a lightweight PAM interface integrated into the memory controller, as shown in \fig\ref{fig:pam-interface}(d). This interface eliminates repeated CPU involvement by offloading layout transformation and inter-device data movement to hardware. It consists of three key components: (1) a command reorder unit at the sender side, (2) an address generation unit at the receiver side, and (3) a dual-port re-layout buffer shared between them. During data transfer, the command reorder unit organizes memory accesses based on device timing and KV token locations, streaming tokens into the re-layout buffer. Once a contiguous region is assembled, data is transmitted to the target device, where the address generation unit issues parallel write addresses. \update{This hardware-managed path enables format-aware KV token migration. Notably, the combination of re-layout and address generation ensures that data is stored in the expected format for each target device, regardless of the internal data width or address mapping differences. This allows PAM to support heterogeneous memory devices such as HBM and DDR, each with distinct column sizes and address mapping schemes. By removing CPU from the critical path, PAM reduces inter-device transfer latency by more than 20 $\times$} and significantly improves scalability by abstracting away layout heterogeneity.

\subsection{Inter-device KV Scheduling}\label{sec:kv-scheduling}

Besides achieving load balance in each PIM device, it also requires dynamic migration of KV tokens across memory hierarchy. However, existing data migration strategies, such as FIFO~\cite{yang2022cachesack, yang2023fifo} and LRU~\cite{lru-management, zahran2007cache, gao2024cost}, fail to handle the dynamics, due to their inability to account for the varying contributions of KV tokens. To address this, we introduce a novel importance factor for each KV token, guiding online scheduling and placement across memory tiers to achieve balanced computation.

\subsubsection{Evaluating KV Importance}

Existing KV sparsity methods~\cite{yang2024post, lee2024infinigen} provide a performance score $\scriptsize S_i^{(j)}$ to indicate the contribution of each KV token $i$ in the $j$-th decoding process. However, scores can vary significantly between adjacent steps (e.g., $\scriptsize S_{i}^{(j-1)}$ and $\scriptsize S_{i}^{(j)}$) ~\cite{cai2024pyramidkv}. As a consequence, relying only on the performance score causes massive data migration, which is prohibitively expensive due to the precious inter-device bandwidth. Instead, PAM introduces a novel importance factor $\scriptsize I_{i}^{(j)}$ as follows. 

\begin{equation}
    I_{i}^{(j)} = \lambda S_{i}^{(j)} + (1-\lambda)I_{i}^{(j-1)}
\label{eq:importance-factor}
\end{equation}

In our evaluation, $\lambda$ is set as $0.6$, a hand-crafted value that achieves the best performance in most cases. Instead of using only the current performance score $S_{i}^{(j)}$, PAM also considers the temporal importance factor $I_{i}^{(j-1)}$ of the previous step to estimate the significance of each KV token. This enables slight adjustments on the importance factors during decoding. Then, we further evaluate the importance of all KV tokens stored on a particular device:

\begin{equation}
    IS^{(j)}_{\mathbb{D}} = \frac{\sum_{i, i\in \mathbb{D}}I_{i}^{(j)}}{\# KV \ tokens}
\label{eq:device-importance}
\end{equation}

where $IS^j_{\mathbb{D}}$ is the cumulative importance score of all KV tokens ($\# KV tokens$) stored on device $\mathbb{D}$ at step $j$. This enables more informed load-balancing decisions across heterogeneous PIM devices.

\subsubsection{Greedy Scheduling Algorithm}

The KV scheduling strategy aims to balance the computation load across heterogeneous devices by adjusting the stored locations of KV tokens, based on our predefined importance ratios of balance:
\begin{equation}
    IS^{(j)}_{{H}} : IS^{(j)}_{{D}} : IS^{(j)}_{{S}}  = x : y : 1
\end{equation}

where $IS_{H}, IS_{D}$ and $IS_{S}$ indicate the importance score of HBM, DDR and SSD, respectively. The predefined ratios $x$ and $y$, obtained via offline profiling, remain fixed during execution. We explore various $x,y$ configurations across datasets to identify the optimal settings for cross-device load balancing. 
\update{Our results show that parameters $x$ and $y$ are architecture-dependent but workload-agnostic, thanks to the attention sparsity algorithm~\cite{ribar2023sparq, yang2024post}. This approach normalizes token scores across datasets, eliminating the need for workload-specific tuning. Thus, $x$ and $y$ only require adjustment when deploying new LLM architectures.}

Bearing the goals of fast and minimal-data-movement load-balancing, we propose a simple yet effective greedy online scheduling strategy. As shown in Algorithm \ref{alg:kv-scheduling}, we adjust the locations of KV tokens based on the predefined ratios with two steps. 

\IncMargin{-0.2em}
\begin{algorithm} 
\small
\caption{Online KV Scheduling}
\label{alg:kv-scheduling} 
\SetKwInOut{Input}{Input}\SetKwInOut{Output}{Output}

\Input{Importance Score $IS^{(j)}_{\mathbb{H}}, IS^{(j)}_{\mathbb{D}}, IS^{(j)}_{\mathbb{S}}$\; \quad \quad \quad \quad Predetermined Ratio $x,y$;} 
\Output{Updated KV Management Table;}
 \BlankLine 
 $x^{*} = \frac{IS^{(j)}_{\mathbb{H}}}{IS^{(j)}_{\mathbb{S}}}$, $y^{*} = \frac{ IS^{(j)}_{\mathbb{D}}}{IS^{(j)}_{\mathbb{S}}}$; \hfill $\rhd$  \textcolor{magenta}{Compute the corresponding ratio}\;
 \While{$(x^{*} + y^{*}) < (x+y)$}{
    Find the least importance KV token in DDR $\mathbb{D}_r$\; 
    Find the most important KV token in SSD $\mathbb{S}_t$\; 
    Swap the corresponding KV token $\mathbb{D}_r \leftrightarrow \mathbb{S}_t$\;
    Update $x^{*}, y^{*}$\;
 }

 \While{$\frac{x^{*}}{y^{*}} < \frac{x}{y}$}{
    Find the least importance KV token in HBM $\mathbb{H}_r$\; 
    Find the most important KV token in DDR $\mathbb{D}_t$\; 
    Swap the corresponding KV token $\mathbb{H}_r \leftrightarrow \mathbb{D}_t$\;
    Update $x^{*}, y^{*}$\;
 }

\end{algorithm}

We initially treat HBM and DDR as a unified device and check if the current KV token distribution matches the target ratio. If low-importance tokens exceed the threshold in DDR, we swap the least important DDR tokens with the most important ones in SSD (lines 3–6). After balancing SSD, the strategy manages KV cache between HBM and DDR similarly (lines 8–11), adjusting only when higher-bandwidth devices hold lower-importance tokens, since low-cost memory is bandwidth-limited. This greedy approach reduces token movement overhead while enabling real-time scheduling.
\update{Specifically, our evaluation demonstrates that only 0.7\% of the total KV tokens require adjustment, with SSD-to-DDR data transfers accounting for less than 0.1\% in each decoding step. However, since important-token divergence accumulates over steps (see Figure \ref{fig:locality}), it is still necessary to adjust inter-device KV placement (evaluation in \textsection \ref{sec:ablation}). Consequently, the lightweight scheduler minimizes unnecessary data movement and ensures a load-balanced processing across PIM tiers, at negligible runtime cost.}

\section{Evaluation}

\subsection{Evaluation Methodology}\label{sec:experimental-setup}

\textbf{Baseline Systems.}
We compare PAM against four systems (vLLM-offloading, AttAcc!, L-PIM and LS-PIM) on a heterogeneous platform (8$\times$H100-80GB GPUs, 1280GB DDR4, 8TB SSD), comparable to DGX-H100~\cite{dgx-h100} and EC2 p5.48xlarge~\cite{amazon-e2-p5}. 


\textbf{vLLM-offloading} (GPU-only): vLLM~\cite{vllm-project} is a high-throughput serving engine managed with paged attention~\cite{kwon2023efficient}, optimized with FlashAttention and FlashDecoding~\cite{dao2022flashattention, flashdecoding}. We extend it with a hierarchical memory design, adopting DeepSpeed-Inference~\cite{aminabadi2022deepspeed}’s offloading to satisfy the memory demands of large-scaling serving.

\textbf{AttAcc!} (GPU+HBM-PIM): AttAcc!~\cite{park2024attacc} integrates HBM-PIM to boost GPU cluster performance by offloading attention while assigning FC operations to GPUs. It does not support KV cache offloading to host memory or SSD.

\textbf{L-PIM \& LS-PIM} (GPU+Layered PIM): To assess PAM's benefits, we include L-PIM and LS-PIM, which use layered PIM in multi-GPU setups. L-PIM mimics AttAcc!, prioritizing KV cache in higher memory tiers. LS-PIM further utilizes KV sparsity to reduce compute and loading overhead. Neither adopts our PAMattention mechanism or KV management strategies.

\noindent
\textbf{Simulation.}
We built an in-house simulator to evaluate PAM and baseline systems, avoiding cross-platform discrepancies consistently. It integrates multiple open-source tools, including LLMServingSim~\cite{cho2024llmservingsim}, LLMCompass~\cite{zhang2024llmcompass}, Ramulator 2.0~\cite{luo2023ramulator, ramulator2.0}, and OpenSSD~\cite{openssd}, with several key enhancements:
\update{1) GPU Simulation: LLMCompass models operators' performance that run on GPUs.
2) KV Transfer \& Computation: Ramulator and OpenSSD are extended to simulate KV token transfers and PIM processing across the memory hierarchy.
3) Interaction and System Scheduling: LLMServingSim is modified to simulate data interaction between GPUs and tiered-PIMs, as well as achieving memory management and online scheduling for PAM system. The evaluation results are cross-verified with the vLLM-based system and AttAcc! simulator. }
These changes enable accurate performance simulation across diverse hardware configurations. We use 4-way tensor and 2-way pipeline parallelism as a representative setup. 



\begin{table}
    \centering
    \caption{Configuration details of PAM.}
    \vspace{-0.3cm}
    \resizebox{\linewidth}{!}{
    \begin{tabular}{c|c|c}
    \hline
    \multicolumn{3}{c}{\textbf{HBM Parameters}} \\
    \hline
     \multicolumn{3}{c}{HBM3\_5.2Gbps, 16GB/HBM$\times$40, 16 channel/HBM}\\
     \multicolumn{3}{c}{2 pseudo channel/ch, 2 rank/pch, 4 bank groups/rank, 4 bank/BG}\\
     \hline
    \multicolumn{3}{c}{\textbf{HBM Timing}} \\
    \hline
        \multicolumn{3}{c}{tRP=19, tRCD=19, tRAS=45, tRRD\_L=4, tBL=4}\\
     \multicolumn{3}{c}{tCCD\_S=2, tCCD\_L=4, tWR=21, tREFI=5070, tFAW=39}\\
    \hline
    \multicolumn{3}{c}{\textbf{DDR Parameters}} \\
    \hline
     \multicolumn{3}{c}{DDR4-3200, 32GB/DDR$\times$40, 2 rank/DDR,8 chip/rank, 2 bank groups/chip, 4 bank/BG}\\
     \hline
    \multicolumn{3}{c}{\textbf{DDR Timing}} \\
    \hline
     \multicolumn{3}{c}{tRC=76, tRCD=24, tCL=24, tRP=24, tBL=4}\\
     \multicolumn{3}{c}{tCCD\_S=4, tCCD\_L=8,tRRD\_S=4, tRRD\_L=6, tFAW=26}\\
    \hline
    \multicolumn{3}{c}{\textbf{SSD Parameters \& Timing}} \\   
     \hline
    \multicolumn{3}{c}{ 256GB/channel $\times$ 64, 4 die per channel, page size = 16KB, 2400MT/s, tR = 30us} \\
    \hline
    \end{tabular}
    }
    \label{tab:pamcfg}
\vspace{-0.3cm}
\end{table}

\begin{figure*}[t]
    \centering
    \includegraphics[width=\linewidth]{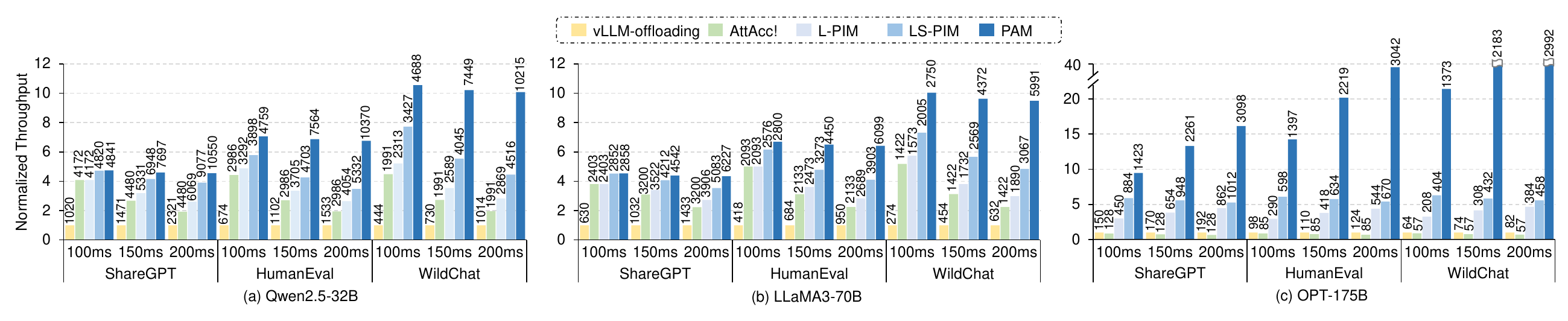}
    \vspace{-0.3cm}
    \caption{The normalized throughput of various SLOs for (a) Qwen2.5-32B (b) LLaMA3-70B (c) OPT-175B. Each number above the bar indicates the maximum batch size per system considering the SLO and capacity limitations.}
    \label{fig:Performance}
\vspace{-0.3cm}
\end{figure*}

\noindent\textbf{Hardware Specifications.}
Table~\ref{tab:pamcfg} outlines the PAM's configuration. Each HBM/DDR4 channel contains 64/128 PIM banks, while each SSD has 8 channels. \update{The operating frequencies of PUs are set to 666MHz (HBM), 500MHz (DDR), and 1GHz (SSD). Power estimates are derived from Ramulator 2.0 and OpenSSD, using voltage and current values from publicly available specifications for HBM3~\cite{park2022192, park2024attacc}, DDR4~\cite{jedec_ddr4, devaux2019true}, and SSD controllers~\cite{FlashtecPCIeGen5NVMeSSDController}. Following prior work~\cite{lee20221ynm, seo2024ianus}, PU power is assumed to be 3$\times$ that of a standard read.}
Under peak power constraints~\cite{park2024attacc,jedec_ddr4,lee2020smartssd}, the per-device compute capacity is 1.6 TFLOPS (HBM), 204 GFLOPS (DDR), and 18 GFLOPS (SSD). These settings are configurable—PAM’s architecture is orthogonal to specific configurations, enabling flexibility.

\noindent\textbf{Evaluation workloads.}
To capture real-world LLM serving scenarios, we evaluate PAM on two workload types: online dialogue tasks and offline long-text processing tasks. 

For \textit{online dialogue tasks}, we utilize three real-world datasets: 1) ShareGPT~\cite{sharegpt} includes a set of conversations scraped from real-world user logs of ChatGPT~\cite{chat-gpt}. 
2) WildChat~\cite{zhao2024wildchat} has a corpus of 1 million real-world user-ChatGPT interactions. 
3) HumanEval~\cite{chen2021evaluating} contains 164 programming problems with a function signature or docstring to evaluate code completion performance. 
\update{These online dialogue tasks usually take a few hundred tokens to complete (the average input/output sequence length is 183/299)}.

For \textit{offline tasks}, we evaluate document summarization using the following datasets: 1) Arxiv\_sum~\cite{cohan2018discourse} is a dataset to estimate the summarization capability for academic papers from arXiv. 
2) Write\_doc~\cite{write-doc} is a benchmark designed to evaluate the performance of writing documents. 
\update{These offline tasks typically involve long-context processing with a sequence length ranging from 1500 $\sim$ 8000, requiring substantial memory capacity.}

\begin{figure}[t]
    \centering
    \includegraphics[width=\linewidth]{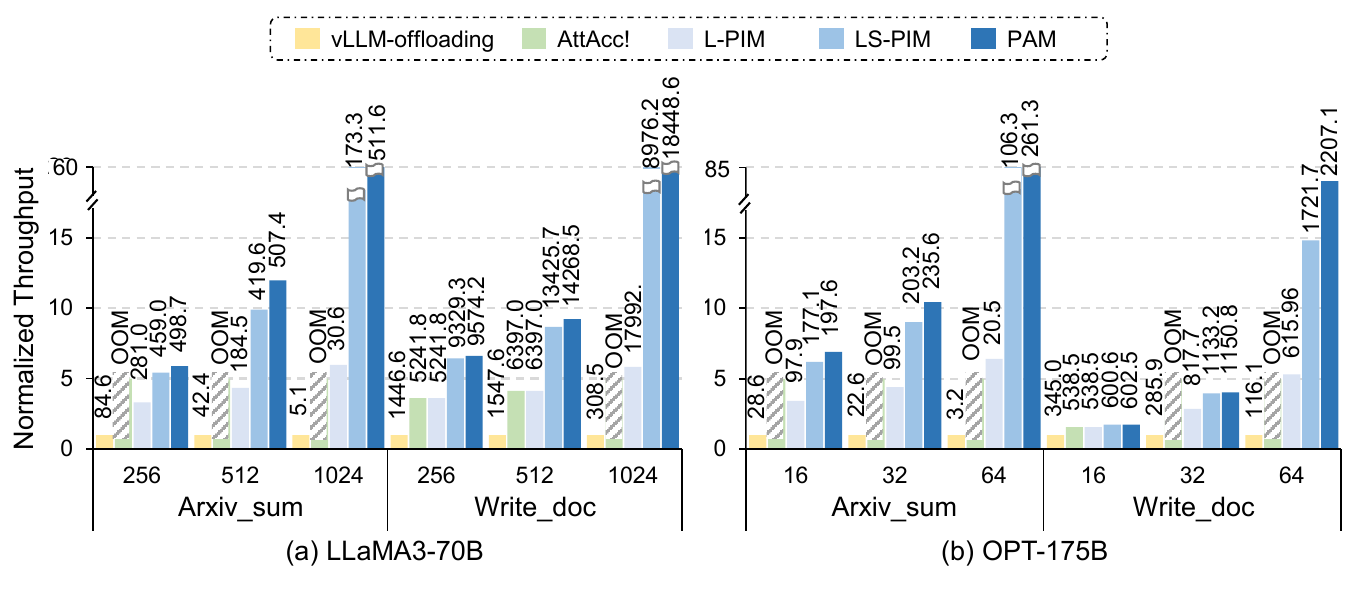}
    \vspace{-0.3cm}
    \caption{The normalized throughput of various batch sizes for offline tasks. Each number above the bar indicates the throughput under the batch size. OOM means out of memory.}
    \label{fig:e2ePerformance}
\vspace{-0.3cm}
\end{figure}

\noindent
\textbf{LLMs \& Algorithms.}
We test PAM using three open-source LLMs: Qwen2.5-32B~\cite{qwen2.5}, LLaMA3-70B~\cite{dubey2024llama3}, and OPT-175B~\cite{zhang2022opt}. \update{In our experiments, all models employ 16-bit floating-point precision (BF16/FP16) for computation, with both weights and KV cache stored in 16-bit floating-point format.} Furthermore, all models leverage the state-of-the-art KV sparsity algorithm~\cite{yang2024post} with $8\times$ compression. The compression yields negligible accuracy loss (under 1\%) across all workloads. Importantly, PAM’s KV management is algorithm-agnostic, supporting diverse compression methods.

\noindent
\textbf{Evaluation Metrics.}
For \textit{online serving}, we measure peak throughput under Service Level Objective (SLO) constraints to ensure real-time responsiveness. For \textit{offline tasks}, we report maximum throughput at a fixed batch size, capturing the system’s efficiency in handling large batches and long-context processing.

\subsection{PAM Performance}\label{sec:pam-performance}

\textbf{Online Serving under diverse SLOs.}
For online tasks, we set SLOs at 100 ms, 150 ms, and 200 ms—faster than typical human reading speeds~\cite{terry2022average}. As shown in \fig\ref{fig:Performance}, PAM consistently supports the largest batch sizes across diverse SLOs. Compared to vLLM-offloading, PAM delivers average throughput gains of $7.20\times$ (Qwen2.5-32B), $6.93\times$ (LLaMA3-70B), and $24.53\times$ (OPT-175B).

PAM shows lower normalized throughput gains on Qwen2.5-32B and LLaMA3-70B than on OPT-175B, as smaller models impose less KV cache pressure. Under the given SLO of 100 ms for ShareGPT, vLLM-offloading keeps the KV cache within HBM for both Qwen2.5-32B and LLaMA3-70B, avoiding performance drops. In this case, AttAcc! achieves the same serving performance as L-PIM. In contrast, serving OPT-175B always exceeds HBM capacity, sharply reducing the supported batch size of vLLM-offloading. L-PIM and LS-PIM encounter similar limitations. Fortunately, PAM overcomes this by efficiently leveraging context locality, maintaining strong performance even when KV cache exceeds both HBM and DDR capacity. In particular, PAM achieves an average $4.54\times$ throughput gain over LS-PIM in various cases.

PAM’s performance gains vary across datasets. It achieves an average improvement of $13.76\times$ on WildChat versus $6.12\times$ on ShareGPT compared to vLLM-offloading. \update{This discrepancy stems from WildChat’s longer average context (738 tokens on average vs. 534 in ShareGPT), which increases the proportion of attention computation—the part PAM optimizes effectively. Longer contexts also require more KV cache capacity, which PAM handles efficiently through its hierarchical memory design and KV management.}

\noindent \textbf{Offline End-to-End Throughput.} 
We evaluate PAM on offline tasks using LLaMA3-70B (batch sizes: 256, 512, 1024) and OPT-175B (batch sizes: 16, 32, 64), as shown in Figure~\ref{fig:e2ePerformance}. PAM consistently achieves the highest throughput, with performance gains increasing at larger batch sizes.

For LLaMA3-70B, PAM improves throughput by $39.2\times$ (Arxiv\_sum) and $25.2\times$ (write\_doc) over vLLM-offloading. For OPT-175B, gains reach $33.0\times$ and $8.26\times$, respectively. These improvements arise as large batches increase KV demands, often exceeding HBM/DDR capacity. AttAcc! fails in most cases due to lack of KV offloading support, while vLLM-offloading, L-PIM, and LS-PIM are bottlenecked by DDR/SSD bandwidth. For example, in L-PIM, SSD holds >65\% of KV data but consumes >93\% of compute time. LS-PIM relies on static KV placement, underutilizing HBM.
In contrast, PAM sustains high throughput by combining PAMAttention with dynamic KV scheduling, effectively leveraging context locality and balancing load across memory tiers for long-context workloads.

\begin{figure}[t]
    \centering
    \includegraphics[width=\linewidth]{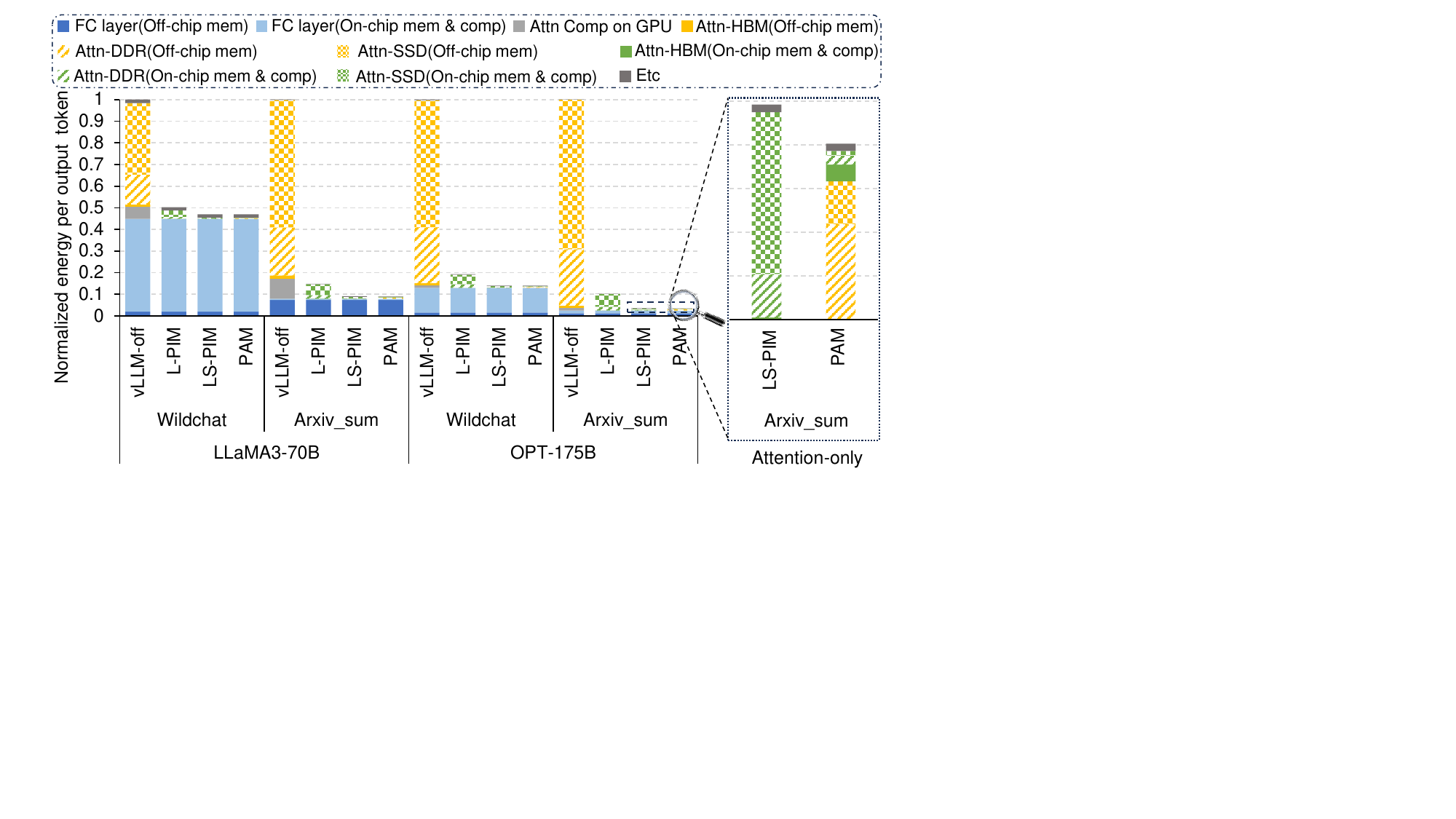}
    \vspace{-0.3cm}
    \caption{The normalized energy per output token for LLaMA3-70B and OPT-175B.}
    \label{fig:energy}
\vspace{-0.3cm}
\end{figure}

In summary, PAM is particularly well-suited for scenarios involving large-scale LLMs, large batch sizes, and long context lengths. Given the widely adopted scaling law~\cite{kaplan2020scaling} and chain-of-thought~\cite{wei2022chain, yao2024tree}, which requires larger-scale LLMs and longer context, PAM provides promising superiority for future LLM serving.

\subsection{Energy Efficiency}\label{sec:energy}
We evaluate PAM’s energy breakdown using LLaMA3-70B and OPT-175B on Wildchat and Arxiv\_sum. For online tasks, batch sizes are 8192 (LLaMA3) and 512 (OPT); for offline, 1024 and 64, respectively.

As shown in \fig\ref{fig:energy}, PAM reduces power by 53.1\%$\sim$92.7\% vs. vLLM-offloading, 7.8\%$\sim$66.9\% vs. L-PIM, and 4.0\%$\sim$18.7\% (attention only) vs. LS-PIM. Gains over vLLM-offloading come from avoiding excessive off-chip KV transfers—e.g., OPT-175B on Arxiv\_sum incurs 2304 GB of KV movement (>95\% energy) in vLLM-offloading, which PAM largely eliminates. Compared to L-PIM, PAM’s sparse attention reduces both compute and memory energy. LS-PIM’s savings are outpaced by PAM’s improved locality and scheduling, which shift more work to energy-efficient HBM. Although PAM incurs inter-device transfers, its tiered RU design and KV scheduling keep overall energy lower, as shown in \fig\ref{fig:energy}.



\subsection{Ablation Study} \label{sec:ablation}

\begin{figure}
    \centering
    \includegraphics[width=\linewidth]{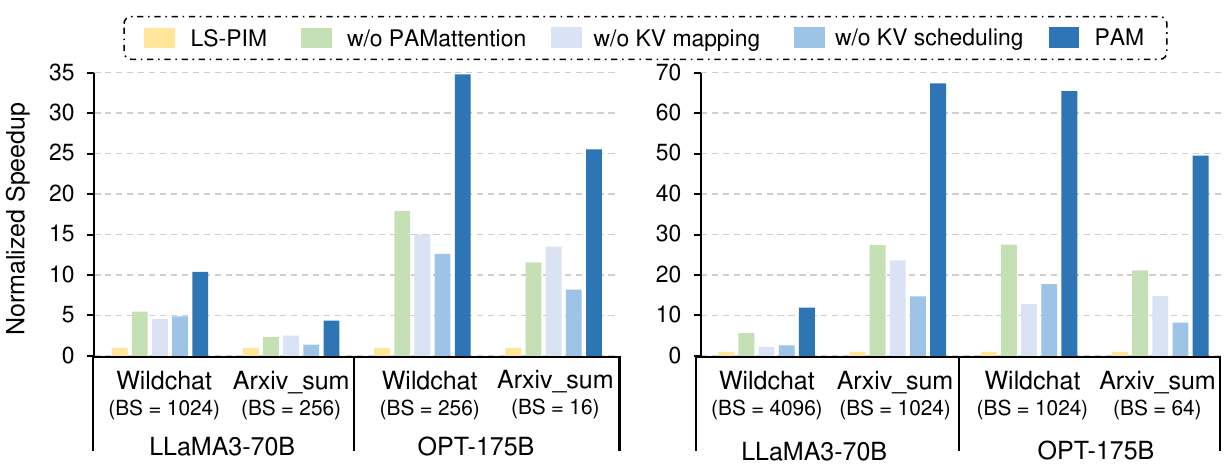}
    \vspace{-0.3cm}
    \caption{\update{Ablation study on proposed PAMattention and KV-centric management strategies.}} 
    \label{fig:ablation}
\vspace{-0.3cm}
\end{figure}

To evaluate the techniques in \textsection\ref{sec:pam-attention} and \textsection\ref{sec:kv-management}, we compare normalized attention computation speedups across LLMs and optimization strategies, using LS-PIM as the baseline (normalized to 1). "w/o PAMattention" uses fixed tiling as in FlashAttention and FlashDecoding~\cite{dao2022flashattention, flashdecoding}; "w/o KV mapping" balances device loads but applies naive KV distribution, ignoring access frequency; and \update{"w/o KV scheduling" disables inter-device scheduling, keeping KV in its original placement.} \fig\ref{fig:ablation} shows results under two batch sizes.


In the small batch setting (left of \fig\ref{fig:ablation}), \update{PAM achieves a $18.7\times$ average speedup over LS-PIM, and $1.93\times$, $2.06\times$, and $2.74\times$ over the "w/o PAMattention," "w/o KV mapping," and "w/o scheduling" variants, respectively}. In large batches (right), PAM's gains rise to $48.56\times$ over LS-PIM, and \update{$2.35\times$, $4.15\times$, and $4.62\times$ over the baselines.} These improvements stem from PAM’s ability to exploit context locality and mitigate bottlenecks in low-bandwidth devices (e.g., SSDs), where LS-PIM suffers from KV overload. \update{Without KV scheduling, cross-tier load balancing remains ineffective. Specifically, while only a small number of KV tokens migrate per decoding step, significant divergence in important KV tokens accumulates over multiple steps. Consequently, the absence of KV scheduling severely impacts attention computation performance, particularly in long-context generation scenarios.} PAM overcomes this via intra-device mapping and inter-device scheduling. Additionally, PAMattention consistently improves performance across models due to shared LLM attention structures, with larger KV caches amplifying the benefits of mapping and scheduling.

\subsection{Area Cost Evaluation}\label{sec:cost}

\begin{table}[t]
\centering
\caption{The additional area of PAM in the memory hierarchy.}
\vspace{-0.3cm}
\label{tab:area_overhead}
\resizebox{\linewidth}{!}{
\begin{tabular}{c|ccc}
\toprule
\multicolumn{2}{c}{\textbf{Components}} & \textbf{Area ($mm^2$)} & \textbf{Area Ratio} \\ 
\cmidrule(lr){1-2}\cmidrule(lr){3-3}\cmidrule(lr){4-4}
\multirow{2}{*}{HBM-PIM} & PU \& RU & $11.2 mm^2$ & 9.25\% (of dram die \cite{park2022192}) \\ 
& PAM interface & $0.85 mm^2$ & 0.703\% (of logic die)\\ 
\midrule
\multirow{2}{*}{\shortstack{DDR-PIM}} & PU \& RU & $5.04 mm^2$ & 5.04\% (of dram chip \cite{samsung_ddr4_2018})\\ 
& PAM interface & $0.85 mm^2$ & 0.750\% (of center buffer) \\ 
\midrule
\multirow{2}{*}{\shortstack{SSD-PIM \\ Controller}} & PU \& RU & $8.16 mm^2$ & 7.592\% (of Controller \cite{TC2200PCIeGen4NVMeDRAMlessSSDController})\\ 
& PAM interface & $0.85 mm^2$ & 0.578\% (of Controller)\\ 
\bottomrule
\end{tabular}}
\vspace{-0.3cm}
\end{table}

\update{The area overhead of PAM mainly arises from integrating Processing Units (PUs), Reduction Units (RUs), and PAM interface logic. We implemented these modules in Verilog and synthesized them using Synopsys Design Compiler~\cite{synopsys.org}. For HBM-PIM and DDR-PIM, PUs and RUs are synthesized in DRAM-compatible 1z-nm and 2x-nm processes~\cite{park2022192, jedec_ddr4, devaux2019true}, which are 10× less dense than logic processes\cite{devaux2019true}. The PAM interface and SSD-side logic (PUs, RUs) are synthesized in TSMC 14nm~\cite{FlashtecPCIeGen5NVMeSSDController, TC2200PCIeGen4NVMeDRAMlessSSDController}.}

\update{For the HBM-PIM configuration, the additional logic occupies 11.2$mm^2$ per DRAM die, 9.25\% of a typical 121$mm^2$ HBM3 die~\cite{park2022192}. Each DRAM die integrates 64 PUs (0.13$mm^2$ each) and 16 RUs (0.18$mm^2$ each). Within each PU, the area is distributed among compute datapaths (67\%), buffer storage (11\%), and control logic (22\%). In the RU, approximately 83\% of the area is dedicated to computing logic.\footnote{\update{Compared to general-purpose designs such as Samsung’s HBM-PIM\cite{lee2021hardware} and UPMEM\cite{devaux2019true}, the additional logic in our design omits complex control units, which contributes to a notable reduction in area cost. Furthermore, for the HBM-PIM case, our evaluation assumes an HBM3-based implementation using a more advanced process node. If synthesized using the same older technology node as Samsung’s HBM-PIM, the total area overhead would increase to 23.4$mm^2$ out of 84.4$mm^2$ ($\sim$ 27.7\%).}}
For the DDR-PIM configuration, each DRAM chip integrates 8 PUs and 2 RUs, contributing a total area overhead of 5.04$mm^2$, or 5.04\% of a typical 100$mm^2$ DDR4 die~\cite{devaux2019true}. The area of each PU and RU is 0.46$mm^2$ and 0.68$mm^2$, respectively.
In the SSD-PIM, the compute extensions—including 64 PUs, 8 RUs, and a shared buffer—occupy an additional 8.16$mm^2$, accounting for 7.59\% of a typical SSD controller area~\cite{TC2200PCIeGen4NVMeDRAMlessSSDController}.
Overall, the added logic is modest and remains within feasible die area budgets under current manufacturing constraints. The overhead is well-justified by the substantial performance gains reported in our evaluation.}



\subsection{\update{Scalability}\label{sec:exp-scalability}}
\begin{figure}
    \centering
    \includegraphics[width=0.85\linewidth]{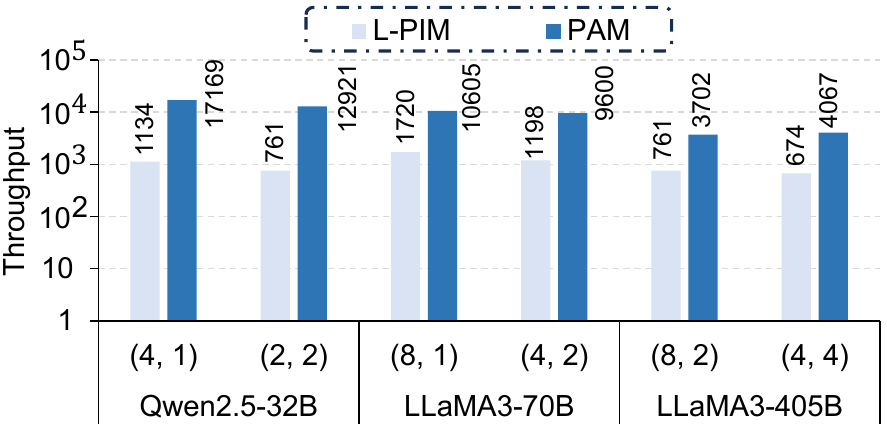}
    \vspace{-0.3cm}
    \caption{\update{Throughput of scaled PAM systems with different numbers of PAM instances under tensor parallelism (TP) and pipeline parallelism (PP), noted as (TP, PP).}} 
    \label{fig:scale}
\vspace{-0.3cm}
\end{figure}

\update{To evaluate PAM's scalability, we ran experiments using the Write\_doc task with 1024 requests, scaling the number of GPUs and storage devices proportionally. The system was extended using standard tensor parallelism (TP) and pipeline parallelism (PP). As shown in Figure~\ref{fig:scale}, PAM consistently outperforms L-PIM, achieving $6.03\times$–$16.96\times$ higher performance across configurations, demonstrating strong scalability.
TP generally yields higher throughput than PP, as PP suffers from pipeline bubbles and added latency. However, TP's communication overhead grows with scale, becoming a key consideration for large deployments.}

\section{Conclusion}\label{sec:conclusion}

We present PAM, an innovative LLM serving system that optimizes costly KV-related operations. PAM integrates processing units within memory devices and introduces a "processing across memory" paradigm to address the high memory bandwidth and capacity demands of LLM serving. Leveraging intrinsic context locality, PAM enables efficient KV allocation across the memory hierarchy. We also propose PAMattention, a novel algorithm using token-wise parallelism to support attention computation across heterogeneous PIM devices. Additionally, PAM implements high-performance KV management through intra-device KV mapping and inter-device KV scheduling, balancing attention computations across the memory hierarchy. Overall, PAM enhances LLM serving efficiency, enabling cost-effective large-scale AI deployment.



\bibliographystyle{ACM-Reference-Format}
\bibliography{refs}

\end{document}